\documentclass[usegraphicx,useAMS,usenatbib]{mn2e}
\usepackage{times}
\usepackage{amssymb}
\usepackage{amsmath}
\usepackage{graphicx}
\usepackage{euscript}
\usepackage{rotating}
\usepackage{epsfig}
\def\mhz{{\rm\thinspace MHz}}
\def\ghz{{\rm\thinspace GHz}}
\def\cmsqps{{\rm\thinspace cm^{2}~s^{-1}}}
\def\kmpspmpc{\hbox{$\rm\thinspace km~s^{-1}~Mpc^{-1}$}}
\def\kev{{\rm\thinspace keV}}
\def\G{{\rm\thinspace G}} %gauss

\begin{document}
%-*-LaTeX-*-
% Copied from gmorris
% Note that some of these call others, eg \kmps uses \km.

% Length
\newcommand{\Mpc}{\rm\thinspace Mpc}
\newcommand{\kpc}{\rm\thinspace kpc}
\newcommand{\pc}{\rm\thinspace pc}
\newcommand{\km}{\rm\thinspace km}
\newcommand{\m}{\rm\thinspace m}
\newcommand{\cm}{\rm\thinspace cm}
\newcommand{\cmps}{\hbox{$\cm\s^{-1}\,$}}
\newcommand{\cmpssq}{\hbox{$\cm\s^{-2}\,$}}
\newcommand{\cmsq}{\hbox{$\cm^2\,$}}
\newcommand{\cmcu}{\hbox{$\cm^3\,$}}
\newcommand{\pcmcu}{\hbox{$\cm^{-3}\,$}}
\newcommand{\pcmcuK}{\hbox{$\cm^{-3}\K\,$}}
% Length

% Time
\newcommand{\yr}{\rm\thinspace yr}
\newcommand{\gyr}{\rm\thinspace Gyr}
\newcommand{\s}{\rm\thinspace s}
\newcommand{\ks}{\rm\thinspace ks}
% Time

% Frequency
\newcommand{\GHz}{\rm\thinspace GHz}
\newcommand{\MHz}{\rm\thinspace MHz}
\newcommand{\Hz}{\rm\thinspace Hz}
% Frequency

% Temperature
\newcommand{\K}{\rm\thinspace K}
% Temperature

% Pressure
\newcommand{\Kpcmc}{\hbox{$\K\cm^{-3}\,$}}
% Pressure

% Mass
\newcommand{\g}{\rm\thinspace g}
\newcommand{\gpcm}{\hbox{$\g\cm^{-3}\,$}}
\newcommand{\gpcmps}{\hbox{$\g\cm^{-3}\s^{-1}\,$}}
\newcommand{\gps}{\hbox{$\g\s^{-1}\,$}}
\newcommand{\Msun}{\hbox{$\rm\thinspace M_{\odot}$}}
\newcommand{\Msunpc}{\hbox{$\Msun\pc^{-3}\,$}}
\newcommand{\Msunpkpc}{\hbox{$\Msun\kpc^{-1}\,$}}
\newcommand{\Msunppc}{\hbox{$\Msun\pc^{-3}\,$}}
\newcommand{\Msunppcpyr}{\hbox{$\Msun\pc^{-3}\yr^{-1}\,$}}
\newcommand{\Msunpyr}{\hbox{$\Msun\yr^{-1}\,$}}
% Mass

% Energy
\newcommand{\MeV}{\rm\thinspace MeV}
\newcommand{\keV}{\rm\thinspace keV}
\newcommand{\eV}{\rm\thinspace eV}
\newcommand{\erg}{\rm\thinspace erg}
\newcommand{\Jy}{\rm Jy}
\newcommand{\ergpcmc}{\hbox{$\erg\cm^{-3}\,$}}
\newcommand{\ergcmcups}{\hbox{$\erg\cm^3\ps\,$}}
\newcommand{\ergpcmps}{\hbox{$\erg\cm^{-3}\s^{-1}\,$}}
\newcommand{\ergpcmsqps}{\hbox{$\erg\cm^{-2}\s^{-1}\,$}}
\newcommand{\ergpcmsqpspA}{\hbox{$\erg\cm^{-2}\s^{-1}$\AA$^{-1}\,$}}
\newcommand{\ergpcmsqpspsr}{\hbox{$\erg\cm^{-2}\s^{-1}\sr^{-1}\,$}}
\newcommand{\ergpcmcups}{\hbox{$\erg\cm^{-3}\s^{-1}\,$}}
\newcommand{\ergps}{\hbox{$\erg\s^{-1}\,$}}
\newcommand{\ergpspmp}{\hbox{$\erg\s^{-1}\Mpc^{-3}\,$}}
\newcommand{\keVpcmsqpspsr}{\hbox{$\keV\cm^{-2}\s^{-1}\sr^{-1}\,$}}
% Energy

% Force
\newcommand{\dyn}{\rm\thinspace dyn}
\newcommand{\dynpcmsq}{\hbox{$\dyn\cm^{-2}\,$}}
% Force

% Speed
\newcommand{\kmps}{\hbox{$\km\s^{-1}\,$}}
\newcommand{\kmpspmp}{\hbox{$\km\s^{-1}\Mpc{-1}\,$}}
\newcommand{\kmpspMpc}{\hbox{$\kmps\Mpc^{-1}$}}
% Speed

% Luminosity
\newcommand{\Lsun}{\hbox{$\rm\thinspace L_{\odot}$}}
\newcommand{\Lsunppc}{\hbox{$\Lsun\pc^{-3}\,$}}
% Luminosity

% Misc
\newcommand{\Zsun}{\hbox{$\rm\thinspace Z_{\odot}$}}
\newcommand{\gauss}{\rm\thinspace gauss}
\newcommand{\arcsecond}{\rm\thinspace arcsec}
\newcommand{\chisq}{\hbox{$\chi^2$}}
\newcommand{\delchi}{\hbox{$\Delta\chi$}}
\newcommand{\ph}{\rm\thinspace ph}
\newcommand{\sr}{\rm\thinspace sr}
% Misc

% Per something
\newcommand{\pccm}{\hbox{$\cm^{-3}\,$}}
\newcommand{\psqcm}{\hbox{$\cm^{-2}\,$}}
\newcommand{\pcmsq}{\hbox{$\cm^{-2}\,$}}
\newcommand{\pmpc}{\hbox{$\Mpc^{-1}\,$}}
\newcommand{\pmpccu}{\hbox{$\Mpc^{-3}\,$}}
\newcommand{\ps}{\hbox{$\s^{-1}\,$}}
\newcommand{\pHz}{\hbox{$\Hz^{-1}\,$}}
\newcommand{\pcmK}{\hbox{$\cm^{-3}\K$}}
\newcommand{\phpcmsqps}{\hbox{$\ph\cm^{-2}\s^{-1}\,$}}
\newcommand{\psr}{\hbox{$\sr^{-1}\,$}}
% Per something
\newcommand{\pspsqas}{\hbox{$\s^{-1}\,\arcsecond^{-2}\,$}}

\newcommand{\ergpspcmpK}{\hbox{$\erg\s^{-1}\cm^{-1}\K^{-1}\,$}}

\title{Radio Bubbles in Clusters of Galaxies}\author[Dunn, Fabian \& Taylor]
{\parbox[]{6.in} {R.J.H. Dunn$^1$\thanks{E-mail:
      rjhd2@ast.cam.ac.uk}, A.C. Fabian$^1$ and G.B. Taylor$^{2, 3}$\\
    \footnotesize
    $^1$Institute of Astronomy, Madingley Road, Cambridge CB3 0HA, UK\\
    $^2$National Radio Astronomy Observatory, Socorro, NM
    87801, USA\\
    $^3$Kavli Institute of Particle Physics and Cosmology, Menlo Park,
    CA 94025, USA\\
  }}
\maketitle

\begin{abstract}
We extend our earlier work on cluster cores with distinct radio bubbles,
adding more active bubbles, i.e. those with $\ghz$ radio emission, to our sample, and also investigating
``ghost bubbles,'' i.e. those without $\ghz$ radio emission. We have
determined $k$, which is the ratio of the total particle energy  to
that of the electrons radiating between $10 \mhz$ and $10 \ghz$.
Constraints on the ages of the active bubbles confirm that the ratio of the
energy factor, $k$, to the volume filling factor, $f$ lies within the
range $1 \lesssim k/f \lesssim 1000$.  In the assumption that there is
pressure equilibrium between the radio-emitting plasma and the
surrounding thermal X-ray gas, none of the radio lobes has
equipartition between the relativistic particles and the magnetic
field.  A Monte-Carlo simulation of the data led to the conclusion
that there are not enough bubbles present in the current sample to be
able to determine the shape of the population.  An analysis of the
ghost bubbles in our sample showed that on the whole they have higher
upper limits on $k/f$ than the active bubbles, especially when
compared to those in the same cluster.  A study of the Brightest 55 cluster sample shows that 17,
possibly 20, clusters required some form of heating as they have a
short central cooling time, $t_{\rm
  cool}\leq3\gyr$, and a large central temperature drop, $T_{\rm centre}/T_{\rm outer}<1/2$. Of these between
12 (70 per cent) and 15 (75 per cent), contain bubbles.  This
indicates that the duty cycle of bubbles is large in such clusters and
that they can play a major role in the heating process.

\end{abstract}


\section{Introduction}

Radio lobes which emit synchrotron radiation contain relativistic electrons and
magnetic fields and tradiationally equipartition (minimum energy) has
been assumed to obtain the pressures of each component \citep{Burbidge},
When the radio source is found in a cluster the radio emitting lobes can
correspond to decrements in the X-ray emission, which have been
interpreted as bubbles in the Intra Cluster Medium (ICM).  This allows the degeneracy to be removed by
measuring the thermal pressure of the ICM and assuming
pressure equilibrium between these bubbles and the surrounding X-ray
gas.  The lack of strong shocks supports the assumption that the
bubbles are not strongly overpressured.  Many moderate power radio
sources in low redshift clusters imaged with {\it Chandra} show
holes in the X-ray emission 
(e.g. Hydra A, \citep{McNamaraHydra00}; Perseus, \citep{ACF_complex_PER00}; A2052,
\citep{Blanton01}; A2199, \citep{JohnstoneA2199}; Centaurus, \citep{SandersCent02}),
the first of which was discovered in the Perseus cluster with
\emph{ROSAT} \citep{Bohringer}.  Recent compilations were given by
\citet{Birzan04} and \citet{DunnFabian04}.

Here we perform a follow up study, extending the sample out to higher
redshifts where possible, and also including some nearby ellipticals
which harbour radio sources and have bubble-like structures observed
in the X-ray emission as well as known cluster ghost bubbles.  We
attempt to include all bubbles for which both the X-ray data available
to us shows clearly evident bubbles and where there is high resolution
radio data.  Following the approach first detailed in
\citet{Celotti02} and developed in \citet{DunnFabian04}, we determine
$k/f$ in the condition of pressure balance, where $k$ is the
ratio of the total relativistic particle energy to that in electrons
emitting synchrotron radiation between $10\mhz$ and $10\ghz$, and $f$ is the
volume filling factor of the relativistic plasma in the bubble.

The bubbles analysed here now include ones which have no current
GHz radio emission -- so-called 'Ghost Lobes' or 'Ghost Bubbles.'  In
some cases these ghost bubbles occur in clusters where there are
active bubbles and so the evolution of the bubbles' $k/f$ within one cluster can
be traced. The distribution of the limits on $k/f$ is also extended
and the suggestion that the underlying population distribution is
bimodal \citep{DunnFabian04} is no longer present in these
extended data; however the wide spread in $k/f$ remains.

In Section \ref{Dataanal} we describe the calculations and assumptions
used to obtain upper limits on $k/f$ and in Section
\ref{results} we give the basic results from this analysis.  The
present sample is combined with that presented in \citet{DunnFabian04}
in Section \ref{discussion} where the Monte-Carlo simulations are
considered along with the effect of the shape of electron
energy spectrum. The results from the analysis of the
ghost bubbles are presented in Section
\ref{ghostbubbles} and an analysis of the number of bubbles
present in clusters is described in Section \ref{bubblenumber}.  Our conclusions are presented in Section
\ref{concls}.  We use $H_0=70\kmpspmpc$ throughout.

\section{Data Analysis}\label{Dataanal}

We follow the analysis presented in \citet{DunnFabian04} and
\citet{Celotti02} to quantify the properties of the particles present
in the bubbles using standard synchrotron theory.  We give some of the
relevant formulae below; for full details of
the calculations see \citet{DunnFabian04}.

The total energy in a bubble which is emitting synchrotron radiation
between $\nu_{1}=10\mhz$ and $\nu_{2}=10\ghz$, with a spectral index $\alpha$
($S(\nu) \propto \nu^{\alpha}$) is given by
\[
 E_{\rm tot} = kE_{\rm e} + Vf\displaystyle\frac{B^{2}}{8\pi} =
 akB^{-3/2} + bf{B^{2}} \erg ,
\]
\noindent
where $E_{\rm e}$ is the energy in relativistic electrons and $V$ is
the bubble volume ($4\pi r_{\rm l} r_{\rm w}^2/3$).  $k$ is the factor
which accounts for additional energy  from particles accompanying
those inferred from the synchrotron emission and electron energy
distribution, and $f$ is the volume filling fraction of the
relativistic plasma.

For simple equipartition between the energy in particles and that in
the bubbles, the magnetic field strength is
\[
 B_{\rm eq} = \Big(\displaystyle\frac{a}{b}\Big)^{2/7}\Big(\displaystyle\frac{k}{f}\Big)^{2/7}\G.
\]
\noindent
Relaxing the condition for equipartition and assuming that there is
pressure equilibrium between the relativistic gas and the thermal ICM
surrounding the bubbles then $k/f$ can be calculated from
\begin{equation} \label{k/f}
\displaystyle \frac{k}{f} = \Big(P_{\rm th} - \frac{B^2}{8\pi}\Big)\frac{3V}{a}B^{3/2},
\end{equation}
\noindent
where the pressure, $P_{\rm th}$, was obtained from temperature and
density profiles of the cluster.

The minimum value for $k/f=1$ which is for an electron-positron plasma
which fills all of the bubble.  A maximum value for $k/f$ can be determined by differentiating
Equation \ref{k/f} with respect to $B$, giving $k/f_{\rm max}$, which is $50$ per cent greater than the equipartition
value.   The field
for $k/f_{\rm max}$ is the limit up to which the $B^2$ in Equation \ref{k/f} term can be
ignored.  Further increases in $B$ causes this term to become dominant
and $k/f$ decreases until it equals one.  At this point the magnetic
field is 1.53 times greater than the magnetic field at $k/f_{\rm max}$, and 1.15
times the equipartition magnetic field.  

Any further increase in the magnetic field and the magnetic pressure
would be such that the bubble would be over pressured, even with
$k=1$.  Hence for pressure equilibrium fewer particles than observed
would be required.  One explanation of this is that the assumptions of
the electron energy spectrum used are wrong, see Section \ref{electenpop}.

% add figure like in ACF & Celotti???????

As there is current $\ghz$ radio emission observed throughout the
bubble, the synchrotron cooling time of the relativistic
electrons can be used to estimate the ages of the bubbles.  

As there are no strong shocks observed in the ICM of these clusters,
the bubbles must expand slower than the sound speed.  Therefore the age of the
bubbles must be greater than $ t_{\rm sound} = 2r_{\rm l}/c_s $ where
$r_{\rm l}$ is the radius of the bubble. 

Detached bubbles have been observed in a variety of clusters, and they
are assumed to rise upwards at their buoyancy velocity, $ v_{\rm b} =
\sqrt{2gV/SC_{\rm D}} $, 
where $S$ is the cross-sectional area of the bubble, $V$ is the
volume, $g = GM(<R_{\rm dist})/R_{\rm dist}^2$ for the bubble (centre) being at $R_{\rm dist}$ from the
cluster core and $C_{\rm D} = 0.75$ is the drag coefficient
\citep{Churazov01}.  Therefore the
age of the bubble can be estimated as $t_{\rm buoy} = R_{\rm
  dist}/v_{\rm b}$, the travel time to their current position.  The age of
the bubble can also be estimated from the time required to
refill the displaced volume as the bubble rises upward
\citep{McNamaraHydra00}, $t_{\rm refill} = 2R_{\rm dist}\sqrt {r/GM(<R_{\rm dist})}$.

The value obtained for the magnetic field from the bubble timescale shows whether the equipartition solution is possible.  The limits
obtained for $k/f$ may be higher than the one predicted for
equipartition as the equipartition condition is determined from the magnetic field strength; Fig 7. of \citet{Celotti02} shows that
it is possible that limits on $k/f$ are larger than the equipartition
value even though the magnetic
field is less than its equipartition value.

From observations of the H$\alpha$ filaments in the Perseus cluster
the flow behind the western ``ghost'' bubble has been assumed to be
laminar \citep{ACF_Halpha_PER03}.  This implies a
Reynolds number of less than 1000, and a value of \mbox{$4
\times 10^{27} {\rm\thinspace cm^{2}~s^{-1}}$} was obtained for the kinematic
viscosity.  The Reynolds number, $Re$, was calculated for each bubble analysed,
assuming the viscosity was the above value, and also a lower bound on
the viscosity was calculated from the limit on the Reynolds number of
1000 assuming that the flow is laminar in each cluster.

\subsection{Ghost Bubbles}

Ghost Bubbles are those which do not have any current GHz radio
emission, the clearest examples being those found in Perseus.  These
were analysed in \citet{DunnFabian04}, but were not included in
any of the conclusions presented there, nor were the results interpreted at any great
length.  In this work a larger number of ghost bubbles have been
analysed, though the numbers of them are still small.

In some cases archival low frequency radio data from the Very Large Array (VLA)
of the NRAO\footnote{The
National Radio Astronomy Observatory is operated by Associated
Universities, Inc., under cooperative agreement with the National
Science Foundation.} with sufficient resolution
has been obtained in order to obtain the radio fluxes of regions
corresponding to these bubbles
at $330\mhz$.  If this has not been possible, then the value
obtained for the region of interest at $\ghz$ frequencies has been
used as an upper limit.  Those sources which have low frequency
measurements are indicated in Table \ref{source prop}.  In A2597
\citep{ClarkeA2597} and the Perseus cluster \citep{Celotti02} spurs of
low-frequency radio emission extend into some of the ghost bubbles.

The Perseus Halo analysed here results from the interpretation of a
high abundance ridge, which corresponds to the edge of the mini-halo
emission, as an ancient bubble \citep{Sanders_PerHalo_04}.  There is
no clear depression in the X-ray emission, but the interpretation that
this high-abundance gas has been pushed out of the centre of the
cluster by a buoyantly rising bubble is plausible.  There is, of
course, the possibility that this feature is not the result of a
bubble and is due to some other phenomenon.

\begin{table*} 
\caption{\scshape \label{source prop}: Source Properties}
\begin{tabular}{l l l l l l l l l l l}
\hline
\hline
Cluster & Lobe$^{(1)}$ & Redshift & $\alpha$ & $R_{\rm
  dist}^{(2)}$
& $r_{\rm l}$ & $r_{\rm w}$ & 
$M_{\rm encl}$ & $kT$ & $n_{\rm e}$ & References\\
& & & & (kpc) & (kpc) &  (kpc) & ($10^{12} M_{\odot})$ &
(keV) & (${\rm cm}^{-3}$)&\\
\hline
\multicolumn{10}{c}{Active Bubbles}\\
\hline
3C401  &N, R   &0.2010&$-1.1\pm0.1$&25.0 &22.0 &14.0 &4.40    &2.9&0.0062 &1\\
       &S, R   &      &            &34.0 &23.0 &13.0 &8.80    &2.9&0.0050 &\\
4C55.16&N, R   &0.24  &$-1.2\pm0.8$&26.5 &11.2 &14.3 & 3.0    &2.8&0.052  & 2, 3,4\\
       &S, R   &      &            &31.9 &17.4 &14.4 & 4.1    &3.0&0.042  & \\
A262   &E, R   &0.016 &$-1.1\pm0.2$& 5.78& 4.93& 2.72& 0.61   &1.2&0.021  &  5,6, 7, 8, 9\\
       &W, R   &      &$-1.2\pm0.2$& 4.75& 4.75& 2.82& 0.50   &1.2&0.021  & \\
A478   &NE, R  &0.088 &$-1.5\pm0.2$& 2.20& 2.20& 2.20& 1.1    &2.0&0.10   &  9,10\\
       &SW, R  &      &            & 3.40& 3.40& 3.40& 1.1    &2.0&0.10   & \\
A1795  &NW, R  &0.063 &$-1.0\pm0.1$& 3.80& 4.10& 3.10& 0.12   &2.7&0.060  & 11, 12, 13, 14 \\
       &S, R   &      &            & 5.20& 4.80& 2.80& 0.20   &2.7&0.060  & \\
A2029  &NW, R  &0.077 &$-1.6\pm0.4$& 9.36& 7.20& 2.16& 0.30   &4.0&0.076  &13, 15, 16,17\\
       &SE, R  &      &            & 9.36& 6.48& 2.88& 0.30   &4.0&0.076  & \\
M87    &E-CJ, R&0.004 &$-1.2\pm0.4$& 1.50& 1.70& 1.30& 0.080  &1.6&0.20   & 13, 18\\
NGC4472&E, R   &0.004 &$-1.2\pm0.1$&4.03 &2.09 &1.55 &0.20    &0.82&0.0590&19,20,21\\
       &W, R   &      &            &3.18 &2.09 &1.71 &0.20    &0.76&0.0800&\\
NGC4636&NE, R  &0.004 &$-1.2\pm0.3$&0.67 &0.54 &0.32 &0.010   &0.60&0.1100& 19,22,23\\
       &SW, R  &      &            &0.65 &0.62 &0.25 &0.010   &0.60&0.1100&\\
\hline
\multicolumn{10}{c}{Ghost Bubbles}\\
\hline
A85                        &N, X   &0.0555&$-1.5\pm0.5$&14.0 & 5.260&7.00& 5.7 &3.0&0.055& 24,25\\
                           &S, X   &      &            &22.0 &6.42 & 8.64& 8.9 &3.0&0.040& \\
A2597                      &NE, X  &0.083 &$-1.0\pm0.4$&21.0 &7.80&7.80& 3.7 &2.2&0.050&  9,13,26,27,28,29\\
                           &SW, X  &      &            &25.0 &12.0 & 7.80& 4.5 &2.3&0.050& \\
\emph{Centaurus} $^{3}$    &N,X    &0.0104&$-2.0\pm0.1$&6.1  & 3.5 &3.0 & 0.4 &1.5& 0.03&  13,30\\
\emph{Perseus Ghost}       &W, X   &0.018 &$-1.5\pm0.7$&28.0 &3.30&14.0 & 2.0 &3.2&0.033&  13,31,32, 33\\
                           &S, X   &      &            &36.0 & 7.00&12.0 & 2.9 &3.6&0.027& \\
\emph{Perseus Halo}        &SW, R  &0.018 &$-1.1\pm0.3$&73.0 &19.0&15.0 & 9.3 &4.7&0.016& 33\\
RBS797                     &W, X   &0.354 &$-1.5\pm0.5$&40.0 &20.0 &20.0 & 4.0 &4.0&0.21& \\
                           &E, X   &       &           &40.0 &20.0 &20.0 & 4.0 &4.0&0.21& 34,35\\
\hline					     	      
\end{tabular}				     
\begin{quote} {\scshape References:}\\
1.  \citet{Reynolds_3C401} 
2.  \citet{Iwasawa4C5516_01};
3.  \citet{Iwasawa4C5516_99};	  
4.  Taylor, G.B., unpublished.;
5.  \citet{Blanton_A262_04}; 
6.  \citet{Parma_A262_86};   
7.  \citet{Fanti_A262_87};   
8.  \citet{White_00};		  
9.  \citet{Reiprich};		  
10. \citet{Sun_A478_03};	  
11.  \citet{Ettori_A1795_02};	  
12.  \citet{Voigt04};		  
13.  \citet{TaylorCent02};   
14.  \citet{Fabian_A1795_01};	  
15.  \citet{Clarke_A2029_04};	  
16.  \citet{TaylorA4059};    
17.  \citet{Lewis_A2029_03}; 
18.  \citet{FormanM87};	  
19.  \citet{Kronawitter}	  
20.  \citet{Ohto_NGC4636}    
21.  \citet{Jones_NGC4636}   
22.  \citet{Ekers_NGC4472}   
23.  \citet{Biller_NGC4472}
24.  \citet{Durret_A85_04}
25.  \citet{Durret_A85_05}	   
26.  \citet{McNamaraA2597};  
27.  \citet{SarazinA2597};   
28.  \citet{ClarkeA2597};
29.  \citet{Pollack_A2597};
30.  \citet{ACF_JSS_Cent05};
31.  \citet{ACF_deep_PER03}; 
32.  \citet{ACF_Halpha_PER03};	  
33.  \citet{Sanders_PerHalo_04};
34.  \citet{Schindler_RBS_01};	  
35.  \citet{DeFilippis_RBS_01}.

{\scshape Notes:}\\
(1) The codes for the Lobes are N---Northern, S---Southern, E---Eastern,
    W---Western etc. , X---sizes from X-ray image, R---sizes from Radio image.  CJ---Counter Jet cavity in M87.\\
(2) All the values given in the above table except the radio power
have an uncertainty associated with them.  Except for
the spectral index, they are not quoted as they have limited effect on
the calculated values.  The effect of the uncertainties in $\alpha$ on
the resultant uncertainties in $k/f$ is
large and so are stated here.  For further discussion see text.\\
(3) The sources in italics have radio data at $330\mhz$.

\end{quote}
\end{table*}

\section{Results}\label{results}

\begin{figure} \centering

\includegraphics[width=1.0 \columnwidth]{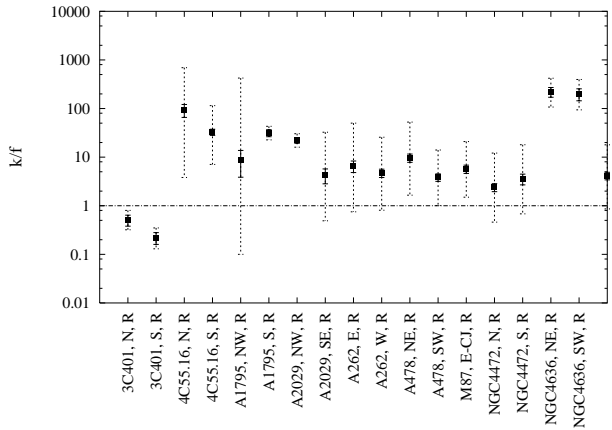}

\caption  {\label{Clustererror} \small{The value of $k/f$ calculated
    from the sound speed limit for each cluster
    analysed, along with the uncertainties arising from the
    uncertainties in $\alpha$ (the dotted bars) and from the
    uncertainties in the other physical parameters of the source (the
    solid bars).  The dotted line shows the minimum value of $k/f$
    possible from the assumptions used in the calculations.}}

\end{figure}

The individual source parameters are listed in Table \ref{source
  prop}, and the resulting values for $k/f$ in Table \ref{kftable}.
The uncertainties presented for the upper limit on the values of $k/f$
arise from uncertainties in the spectral index, $\alpha$, of the radio emission
from the bubble.  As $\alpha$ appears as the exponent in the equation
for the energy in synchrotron emitting elections, any uncertainty in its value has a large effect on the
range of allowed values.  In some cases a value has not been able to
be determined accurately, resulting in the choice of a range which is
large enough to encompass most of the plausible values.  In Fig.
\ref{Clustererror} the solid error bars show the range of values for
the upper limit on $k/f$ which arise from uncertainties in all other
parameters used, and they are on the whole, less than the ones
resulting from uncertainties in $\alpha$ (dotted error bars).

\begin{figure} \centering

\includegraphics[width=1.0 \columnwidth]{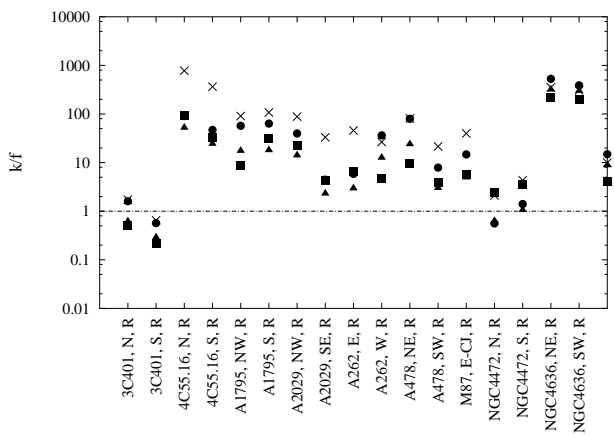}

\caption  {\label{Clusternew} \small{The value of $k/f$ for each cluster
    analysed.  The solid square, circle and triangle symbols denote the $k/f$ values from the sound speed,
  buoyancy and refilling timescales respectively.  The $\times$ symbol
  denotes the equipartition values}}

\end{figure}

In this sample there is good agreement between limits on $k/f$ from different
timescale (Fig. \ref{Clusternew}).  Any disagreement between the
limits arises from the differences in the timescales calculated for
the bubbles (for further discussion see Section \ref{timescales}).  In this
sample only A1795 and A478 had radio images at two frequencies
analysed, and the agreement between the limits obtained at the
different frequencies is very good.  The agreement between
bubbles within one cluster is also very good.

Some limits on $k/f$ fall below the line corresponding to the minimum
possible $k/f$, e.g. 3C401 here and Hydra A in \citet{DunnFabian04}.
The interpretation of $k/f<1$ is that for pressure balance fewer
particles than those observed are required, which could be as a result
of using an inaccurate form of the electron energy spectrum (Section
\ref{electenpop}).  In Hydra A the radio source seems to have blown very large cavities in
the ICM \citep{NulsenHydra04}, and so it may be over-pressured.  As such the assumptions used in the
calculation of the $k/f$ limits may not hold.  In 3C401 although there
is evidence for interaction between the radio source and the ICM
\citep{Reynolds_3C401}, there are no clear cavities in the X-ray
emission, however the central regions of the cluster are not
spherically symmetric, and the bar of X-ray emission could be as a
result of cavities in the cluster.  As such the uncertainties in the limit on $k/f$ are
probably larger than shown.

\subsection{Timescales}\label{timescales}

The differences between the methods for estimating the timescales (ages) of the
lobes are important in this sample.  The sound-speed timescale
is the timescale for the expansion of the bubble at a speed less than
the sound speed of the medium it is observed to be in currently, as there are no
strong shocks observed at the edges of the bubbles.  However it is
possible that the expansion is still slightly supersonic;
\citet{Blanton01} state that in A2052 speeds of up to
$\mathcal{M}a\sim1.2$ are allowed if there are not strong shocks.  Also, as the bubbles are assumed to be created from a
relativistic jet, there would have been some supersonic expansion in
the past.  This timescale estimate is probably the most appropriate
one for the young, powered bubbles.  Some sources, especially those with hotspot emission, e.g. Cygnus A,
are likely to be expanding much faster than the local sound speed.  In
this case the sound speed timescale would be much too small and the
upper limit on $k/f$ would rise significantly.  

The buoyancy timescale is the time taken for a bubble of the observed
size to rise buoyantly to its current position in the cluster assuming
a uniform medium, with properties as observed at its current
location.  However, as the bubble is likely to be expanding as it is
rising, as well as the fact that the medium is non-uniform, means that
there are some uncertainties on this timescale beyond those arising
from the uncertainties in the measured parameters.  This timescale is
probably the most appropriate for the ghost bubbles, which presumably have detached
from the jet that formed them a long time ago and so are
rising buoyantly.  The refilling timescale is also a rise-time-like timescale, as it
gives the time taken to refill the displaced volume.  Some of the ghost bubbles are at a
much larger distance from the centre of the cluster than young ones,
but they are not that much larger in radius.

These differences in the type of timescales calculated can explain the
fact that, for example, the sound speed timescales for the Perseus Ghost bubbles and
the Halo are much less than the other two estimates.  The sound speed
timescale measures the expansion of the ghost bubble at the sound
speed of the medium, which increases as the temperature (and hence
distance from the centre of the cluster) increases.

The form for the synchrotron lifetime is
only valid  for sufficiently large magnetic fields.  Inverse Compton (IC)
losses dominate when the energy density of the Cosmic Microwave
Background exceeds that of the magnetic field, $U_{\rm
  B}=B^2/8\pi$, which corresponds to
$B=B_{\rm CMB}$.  Although all the clusters analysed are in the local
Universe, it was checked whether the inferred magnetic field was
larger than $B_{\rm CMB}$.  The lifetime determined
from this limiting magnetic field for $1\ghz$ electrons ($t_{\rm CMB}$) was also calculated and is tabulated
in the Appendix.  The only cluster where any timescale exceeds the
value of $t_{\rm CMB}$ is RBS797 for the refilling timescale estimate.

\begin{table*} 
\caption{\scshape Physical $k/f$ Values}
\begin{tabular}{l l l l l l l l l l}
\hline
\hline
Cluster & Lobe$^{(1)}$ & Pressure  & $Re$ & Viscosity$^{(2)}$ & Energy$^{(3)}$  & $k/f_{\rm eq}$&
$k/f_{\rm sound}^{(4)}$  & $k/f_{\rm buoyancy}$
&$k/f_{\rm refill}$\\
& & $(\rm eV/cm^3)$  & & $(10^{27}\!\cmsqps)$ & $(10^{58}\erg)$  &  &   &  &\\
\hline
\multicolumn{10}{c}{Active Bubbles}\\
\hline
3C401  &N, R   &39.6 &1938   &7.75  &3.36   &1.72  & $  0.51 \;^{ 0.80 }_{ 0.32}$ &$ 1.60  \;^{  2.51 }_{ 1.00 }$ &$ 0.62 \;^{ 0.98  }_{ 0.39 }$ \\
       &S, R   &31.9 &1990   &7.96  &2.44   &0.65  & $   0.22\;^{ 0.35}_{ 0.13 }$ &$  0.57 \;^{  0.91 }_{ 0.35 }$ &$  0.29\;^{ 0.47 }_{ 0.18 }$ \\
4C55.16&N, R   &  320&   879 & 3.54 & 14.4  &776   & $ 93.5  \;^{692  }_{  3.82}$ &$ 90.4  \;^{ 669   }_{  3.69}$ &$ 52.4 \;^{388   }_{  2.14}$ \\
       &S, R   &  277&  1227 & 4.91 & 19.7  &364   & $ 32.8  \;^{115  }_{  7.08}$ &$ 47.0  \;^{ 165   }_{ 10.2 }$ &$ 24.4 \;^{ 85.7 }_{  5.28}$ \\
A262   &E, R   & 55.4&   301 & 1.20 & 0.040 &352   & $221    \;^{419  }_{107   }$ &$528    \;^{1003   }_{257   }$ &$319   \;^{605   }_{155   }$ \\
       &W, R   & 55.4&   330 & 1.32 & 0.041 &308   & $199    \;^{396  }_{ 94.0 }$ &$387    \;^{ 769   }_{183   }$ &$300   \;^{595   }_{141   }$ \\
A478   &NE, R  &  440&   471 & 1.88 & 0.092 &177   & $ 70.3  \;^{152  }_{ 31.4 }$ &$225    \;^{ 485   }_{100   }$ &$149   \;^{321   }_{ 66.5 }$ \\
       &SW, R  &  440&   585 & 2.34 & 0.34  &653   & $171    \;^{368  }_{ 76.2 }$ &$885    \;^{1909   }_{395   }$ &$307   \;^{661   }_{137   }$ \\
A1795  &NW, R  &  356&  190  & 0.76 & 0.28  &107   & $ 31.6  \;^{ 43.0}_{ 22.6 }$ &$  63.4 \;^{  86.3 }_{  42.5}$ &$ 18.1\ ;^{  24.6}_{  13.0}$ \\
       &S, R   &  356&  191  & 0.76 & 0.26  & 88.4 & $ 22.3  \;^{ 30.4}_{ 16.0 }$ &$  39.8 \;^{  54.2 }_{  28.6}$ &$  14.2\;^{  19.3}_{  10.5}$ \\
A2029  &NW, R  &  669&   153 & 0.61 & 0.44  & 33.1 & $ 4.27  \;^{ 32.5}_{  0.49}$ &$  4.46 \;^{  33.9 }_{  0.51}$ &$  2.31\;^{ 17.6 }_{  0.26}$ \\
       &SE, R  &  669&   170 & 0.68 & 0.71  & 45.8 & $ 6.57  \;^{ 50.0}_{  0.75}$ &$  5.86 \;^{  44.7 }_{  0.67}$ &$  2.96\;^{ 22.5 }_{  0.34}$ \\
M87    &E-CJ, R&  704&   106 & 0.42 & 0.040 & 26.8 & $  8.75 \;^{ 31.0}_{ 1.87 }$ &$ 28.6  \;^{ 102   }_{ 6.13 }$ &$  8.53\;^{ 30.3 }_{  1.83}$ \\
NGC4472&E, R   &106  &83     &0.33  &0.011  &4807  & $   3531\;^{ 5010}_{ 2448 }$ &$ 5143  \;^{ 7296  }_{ 3565 }$ &$ 3209 \;^{ 4553 }_{  2225 }$ \\
       &W, R   &134  &115    &0.45  &0.016  &8727  & $   5323\;^{ 7552}_{  3690}$ &$ 11401 \;^{ 16175 }_{ 7903 }$ &$ 5998 \;^{ 8509 }_{ 4158 }$ \\
NGC4636&NE, R  &145  &12.6   &0.051 &$1.6\!\times\!10^{-4}$&61.5& $ 88.4\;^{ 237}_{ 28.3 }$ &$ 53.2 \;^{ 142 }_{ 17.0 }$ &$ 77.7 \;^{ 208  }_{ 24.9 }$ \\
       &SW, R  &145  &12.4   &0.050 &$1.1\!\times\!10^{-4}$&71.0& $ 95.9\;^{ 257}_{  30.7}$ &$  19.4\;^{ 52.0 }_{  6.22 }$ &$ 94.4\;^{ 253  }_{ 30.2 }$ \\

\hline
\multicolumn{10}{c}{Ghost Bubbles}\\
\hline
A85           &N, X  &  363& 758 & 3.03 & 1.86 &  19462& $ 4631\;^{28513}_{ 581}$&$ 6781\;^{41756}_{  851}$&$ 4390\;^{27030}_{  551}$ \\
              &S, X  &  264& 802 & 3.21 & 2.46 &  40635& $10154\;^{62521}_{1275}$&$ 9936\;^{61178}_{ 1247}$&$ 8315\;^{51201}_{ 1044}$ \\
A2597         &NE, X  &  242& 604 & 2.41 & 2.3 &  268109& $50559\;^{235504}_{7361}$&$ 55871\;^{260247}_{8134}$&$ 40104\;^{186808}_{5839}$ \\
              &SW, X  &  253& 801 & 3.20 & 3.6 &  445843& $54361\;^{253218}_{7914}$&$ 86847\;^{404538}_{12644}$&$ 55729\;^{259589}_{8113}$ \\
Centaurus     &N,X    &   99&58.6 & 0.23 &0.062&   152.7& $ 97.5\;^{ 128.8}_{73.4}$&$ 50.7 \;^{  67.0}_{38.2}$&$  25.1\;^{  33.2}_{18.9}$\\
Perseus Ghost &W, X   &  232& 242 & 9.70 & 3.0 &   16522& $ 8771\;^{ 34615}_{1294}$&$   976\;^{  3853}_{ 144}$&$  1355\;^{  5347}_{ 200}$ \\
              &S, X   &  206& 381 & 1.52 & 4.1 &   17254& $ 5174\;^{ 20419}_{ 763}$&$  1176\;^{  4641}_{ 173}$&$  1339\;^{  5285}_{ 198}$ \\
Perseus Halo  &SW, R   &  172& 894 & 3.58 & 14.5&   49178& $ 7214\;^{  7397}_{5859}$&$  2744\;^{  2814}_{2229}$&$  3024\;^{  3101}_{2456}$ \\
RBS797        &W, X   & 1848& 1353& 5.41 & 29.2&  920472& $20118\;^{126028}_{2482}$&$ 19438\;^{121768}_{2398}$&$ 10313\;^{ 64603}_{1272}$ \\
              &E, X   & 1848& 1353& 5.41 & 29.2&  920472& $20118\;^{126028}_{2482}$&$ 19438\;^{121768}_{2398}$&$ 10313\;^{ 64603}_{1272}$ \\
\hline	

\end{tabular} \label{kftable}
\begin{quote}
{\scshape Notes:} \\
(1) The codes for the Lobes are N---Northern, S---Southern, E---Eastern,
    W---Western etc., X---sizes from X-ray image, R---sizes from Radio image.  CJ---Counter Jet cavity in M87.\\

(2) The viscosity is estimated assuming that the flow is laminar and
has a Reynolds number of $1000$

(3) The energy quoted here is $E=PV$, so the values have to be
    multiplied by the appropriate $\gamma/(\gamma-1)$.\\

(4) The range on the limits on $k/f$ from the uncertainty in the spectral
    index are given by the maximum values (superscript) and minimum
    values (subscript).  The uncertainties from other parameters are
    shown in Fig. \ref{Clustererror}.\\

\end{quote}
\end{table*}

\subsection{NGC4472}

NGC 4472 is an elliptical galaxy in the Virgo Cluster.  As can be seen
from Fig. \ref{Clustererror} the $k/f$ value is much higher (by a
factor of 10) than the
other new clusters, and from Fig. \ref{Clustererrorall} is much
higher than all of the clusters in the sample.  The radio emission
seen from the radio source at the centre of NGC 4472 is only just
above the background, and as such the bubbles may be more akin to
Ghost rather than Active ones.  The results for this galaxy are still
included with the active bubbles in the subsequent analysis as there
is still current $\ghz$ radio emission from the lobes.

\section{Discussion}\label{discussion}

As can be seen in Fig. \ref{Clustererror}, the spread in
the upper limits on $k/f$ reported in \citet{DunnFabian04} is also
present in these bubbles.  As in \citet{DunnFabian04}, the limits placed on the magnetic field
during the course of this calculation means that there cannot be
simple equipartition between the particles and the field present in
the bubble.  The magnetic fields estimated from the sound speed
timescale are $0.01-0.6$ times the equipartition values.  If the particles and field present in the bubbles were in equipartition, then the pressure in the
bubbles would be much less than the pressure from the surrounding ICM,
ranging from 1.5 times (Centaurus) to 160 times (Perseus) too low, with
an average of around 20 times.  Radio sources in clusters often appear
distorted, confined and have steep spectra, which argues for
conditions close to pressure equilibrium.  Interactions may have
increased the internal pressure of these sources \citep{Pollack_A2597}.  

Recent studies of radio galaxies which are not in cluster environments
also imply that the radio lobes are not at equipartition.  \citet{Croston3c66B} amongst others find that the pressure of the
lobes are too low if equipartition is assumed, if $f$ and $k\approx1$, the
lobes being under-pressured by $\sim20\times$; \citet{Hardcastle2000}
state that there is little concrete evidence for the assumption
that the radio lobes are near to their minimum pressures.  Inverse
Compton X-ray emission would allow a check as to whether lobes were at
or close to equipartition, but the emission from several sources (e.g. 3C120,
\citealt{Harris3c120}) is inconsistent with being at equipartition.
The presence of the ICM, constraining the expansion of the lobes may
cause the energy in particles to dominate over the energy in the
magnetic field.

\subsection{Total Sample}

\begin{figure} \centering

\includegraphics[width=1.0 \columnwidth]{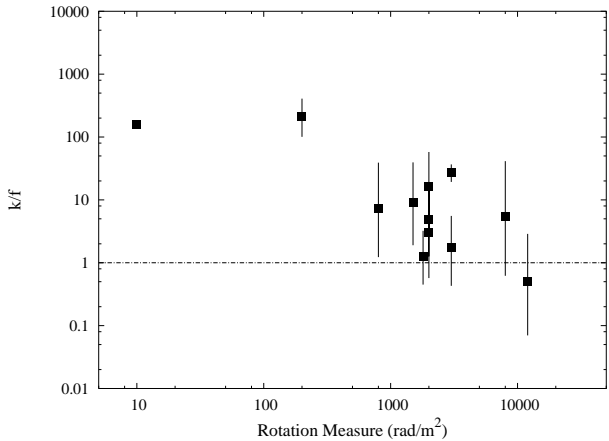}

\caption  {\label{RotMeas} \small{The dependence of $k/f$ on the
    Rotation Measure from the radio source for those in the total
    sample which have a calculated value.  The plot is for
    cluster-averaged $k/f$ to remove double counting.}}

\end{figure}

Combining the data from the sample presented here with that from
\citet{DunnFabian04} allows confirmation that there appears to be no
strong correlation between $k/f$ and any physical parameter of the
cluster or radio source for bubbles from different clusters.  There is
a possibility that the Rotation Measure (RM) of the radio source may give
some indication of $k/f$ (Fig. \ref{RotMeas}).  However not all of the bubbles in the
sample have values for the Rotation Measure of their radio source and
there is a large scatter in $k/f$ for a given Rotation Measure.  The
radio source 3C84 in the Perseus Cluster is completely depolarised and
as such the rotation measure is expected to be large, yet it has a
high $k/f$ so it would not fit the the trend implied by
Fig. \ref{RotMeas}.  

\begin{table}
\caption{\scshape Rotation Measures}
\begin{centering}
\begin{tabular}{l l l}
\hline
\hline
Cluster & Rotation Measure & Reference\\
\hline
A262	 & 200  & 3\\
A1795	 & 3000 & 1\\
A2029	 & 8000 & 1\\
A2052	 & 800  & 1\\
A2199	 & 2000 & 1 \\
A2199I   & 2000 & 1\\
A4059	 & 1500 & 1\\
Centaurus	 & 1800 & 1\\
Cygnus A  & 3000 & 1 \\
Hydra A	 & 12000& 1\\
M84	 & 10   & 2\\
M87	 & 2000 & 1\\
\hline
\end{tabular} \label{rmtable}
\end{centering}
\begin{quote}
{\scshape References:} 1 \citealt{TaylorCent02}; 2
\citealt{LaingBridleM84}; 3 \citealt{Clarke_RotMeas01}
\end{quote}
\end{table}

Recent measurements of the Rotation Measure of A2597 by
\citet{Pollack_A2597} give values for the inner radio source of
$3620\thinspace {\rm rad~m^{-1}}$.  The $k/f$ for the outer ghost bubbles is
$\sim3.5\times 10^4$ which, if the rotation measure were the same for
both, sits far above the other sources in Fig. \ref{RotMeas}.  As the
Rotation Measure of the radio source probes the surrounding ICM, so
higher RMs indicate higher cluster magnetic fields and/or densities.
A higher external magnetic field could contribute to the pressure
acting on the bubbles, but the thermal pressure will dominate.  If the
RM comes from the magnetic field in the ICM, then it will also depend
on the impact parameter through the cluster.  Deeply embedded, new,
bubbles will acquire high RMs and those which have risen up would
acquire lower RMs.  In this case, as the
bubbles rise up through the ICM, their $k/f$ increases (see Section
\ref{ghostbubbles}) and the RM decreases.  This interpretation may be the
explanation for the location of the points in Fig. \ref{RotMeas}, and as such
there is no direct correlation between $k/f$ and RM.  Also, an RM measured in the center of the cluster
towards a relatively young source may not have much to do with an
old outer bubble from the same source.

As there is a large variation in physical parameters between clusters, any correlation that $k/f$
might have with any given parameter could be wiped out by inter-cluster
differences in other parameters.  To try and
remove this effect, the values for the distance of the bubble from the
cluster centre ($R_{\rm dist}$), the timescales, the luminosities and the
temperature of the ICM surrounding the bubbles were scaled using
values for $r_{\rm cool}$, $t_{\rm cool}$, $L(<r_{\rm cool})$ and
$<\!k_{\rm B}T\!>$ from
\citet{Peres1998}, converted to our cosmology.  However, there were no
resultant clear correlations.

\begin{figure} \centering

\includegraphics[width=1.0 \columnwidth]{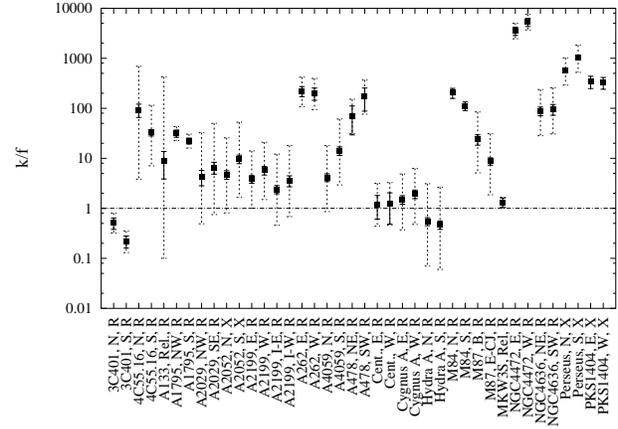}

\caption  {\label{Clustererrorall} \small{The value of $k/f$ calculated
    from the sound speed limit for each cluster
    analysed, along with the uncertainties arising from the
    uncertainties in $\alpha$ (the dotted bars) and from the
    uncertainties in the other physical parameters of the source (the
    solid bars).  The dotted line shows the minimum value of $k/f$
    possible from the assumptions used in the calculations.}}

\end{figure}

\begin{figure} \centering

\includegraphics[width=1.0 \columnwidth]{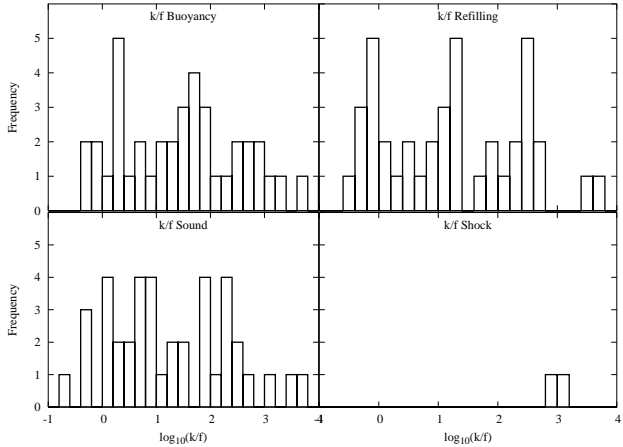}

\caption  {\label{Barchartall} \small{The distribution of the numbers
    of bubbles with given ${\rm log}_{10}(k/f)$ values for the sound
    speed, buoyancy, refilling and shock time-scale calculations.}}

\end{figure}

Fig. \ref{Clustererrorall} shows the $k/f$ limits for all the active
bubbles in the combined sample.  The frequency distribution of $k/f$
is shown in Fig. \ref{Barchartall}.  It can be clearly seen from the
frequency distribution that a single population is the most likely
explanation of the data.  This is contrary to what was presented in
\citet{DunnFabian04}.  The binning presented in Fig.
\ref{Barchartall} does not take into account the uncertainties present
on $k/f$.  

\begin{figure} \centering

\includegraphics[width=1.0 \columnwidth]{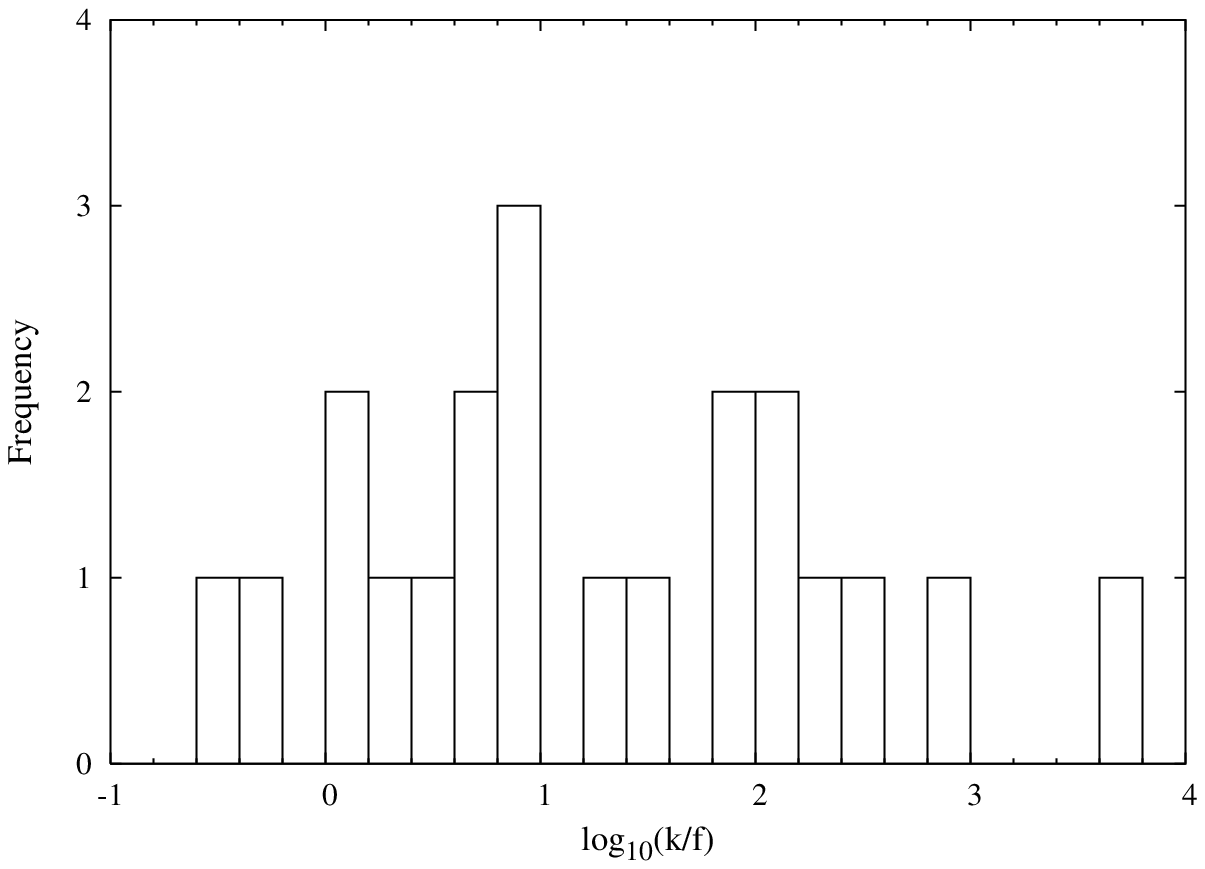}
\includegraphics[width=1.0 \columnwidth]{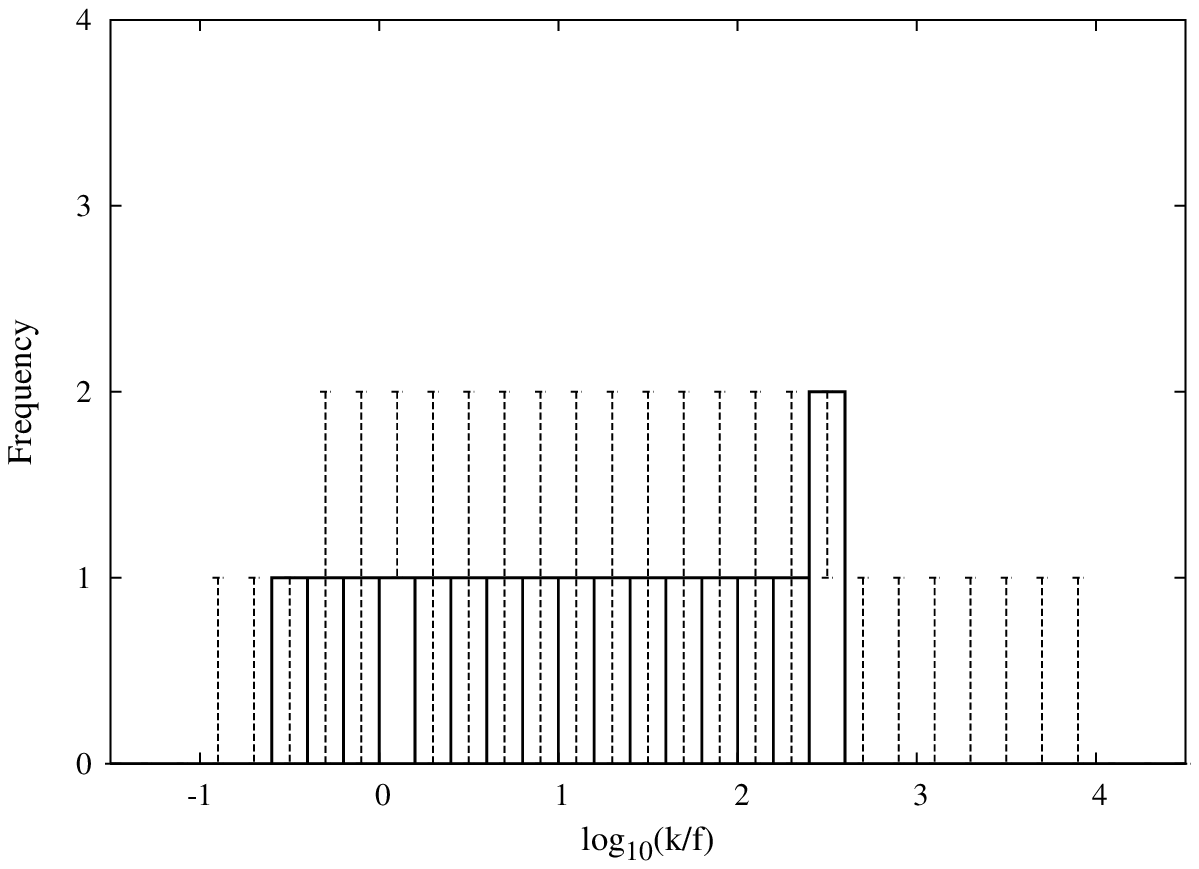}

\caption  {\label{Barcharterror} \small{The distribution of the numbers
    of bubbles with given ${\rm log}_{10}(k/f)$ values for the sound
    speed calculation: {\scshape Top} for cluster averaged values; and
    {\scshape Bottom} the results of the Monte Carlo runs, with the
    uncertainties on the bin values as shown.}}
\end{figure}

For many of the clusters analysed here, there are two bubbles per
cluster, and hence two limits on $k/f$.  For a statistical analysis
this poses a problem as the two limits are not independent as they are
drawn from the same cluster.  To remove this difficulty a
``cluster-average'' was created for each cluster.  If there are two
bubbles per cluster then the two values for all the quantities
calculated during the course of the analysis were averaged.  The
uncertainties on the values were also just averaged.  The resultant data were then
binned up as before.  It was also checked at this stage as to whether
this ``double counting'' could have hidden some correlation, but there
was no such case.  The resulting frequency distribution is shown in
Fig. \ref{Barcharterror} (top).

To determine the form of underlying distribution for $k/f$ a
Monte Carlo simulation of the data was performed to obtain errors on
the frequency distribution.  The resulting frequency distribution is
shown in Fig. \ref{Barcharterror} (bottom).

As can be seen in the figure, there is no clear choice for a
distribution of $k/f$.  There are only 20 clusters (the Inner
bubble pair of A2199 are counted separately from the outer pair,
making 21 in the cluster averaged data) whose values for the limit on
$k/f$ spread over a few orders of magnitude.  This limits the scope of
any conclusions which can be drawn from the distribution, and many, many
more bubbles would need to be added to this sample such that any
underlying distribution could be determined.  All that can be said from
this data is that there is a definite spread in the limits on $k/f$,
which for bubbles with active $\ghz$ radio emission spreads over $\sim
10^3$. 

This large spread in $k/f$ is the result of a signification population of
non-relativistic particles present in some of the bubbles.  The origin
of these particles is unknown; one suggestion is that the jets may intrinsically
contain protons when they are formed.  On the other hand, if the jets
produced by the central black hole are pure
electron-positron, as they travel out into the ICM they may pick up
material (probably thermal protons).  This entrainment is stochastic,
depending on the environment of the jet in the first few kpc around
the radio source.  This varies from cluster to cluster and as such
there is a large range in the values of $k/f$.  It would be expected
that powerful jets would punch through the ICM and not entrain much
material, so ending up with lower $k/f$, for example Hydra A and
Cygnus A; those sources which are less powerful would pick up more
material (given the same ICM).  There are just too many differences in
the physical conditions in clusters which result in there being no correlation with only
one parameter.

We assume that $k$ has most effect on $k/f$ in these bubbles and that
$f$ is not thought to vary much.  \citet{Schmidt2002} analysed the in
Perseus and rule out at the 3 sigma level of gas cooler than 11keV
filling the entire hole, and at the same level, 6.5keV gas filling 1/3
of the hole.  It is unlikely that $f$ will change much during the
evolution of the bubble as even though the electrons age, they are
still relativistic.

\subsection{Low-Energy Electron Population \& Spectral Indices}\label{electenpop}

As was discussed in \citet{DunnFabian04} the spectral index of the
radio emission is the overwhelming uncertainty on the calculated value
of $k/f$.  The value of the spectral index is vital as it is used when
extrapolating the electron energy spectrum from $\ghz$ frequencies
down to $\mhz$ frequencies.  The assumption so far has been that the
spectral index is smooth between $\nu_1=10\mhz$ and $\nu_2=10\ghz$.  However it is
very difficult to obtain observations with sufficient resolution at
these lower frequencies, and so it is uncertain what happens to the
electron population.  

In ghost lobes spectral steepening as a result of spectral ageing is of
importance, especially if there is no re-acceleration.  The spectral
indices in the Perseus ghost lobes, the mini-halo and in Centaurus are
steep, $-1.5 \to -2.0$, suggesting that the electrons have aged in these
regions, implying that there is minimal re-acceleration present
in these lobes \citep{ACF_JSS_Cent05}.  However, this spectral index causes
problems when calculating $k/f$, as extrapolating the electron
population with such a steep index will undoubtedly over-estimate the
low energy electron population (with the assumption that the un-aged
spectrum is $\sim -1.0$).  

Harris (2004) discusses various possibilities for the low end of the
relativistic electron spectrum:
\begin{enumerate}
\item $\alpha$ steepens with decreasing frequency so that
  extrapolation under-estimates the low energy electron population.
\item $\alpha$ remains essentially constant.
\item $\alpha$ flattens towards 0, so that there are fewer low energy
  electrons as are predicted by extrapolation.
\item There is a low energy cut-off so that there are no low energy
  electrons.
\end{enumerate}
We now briefly discuss the implications of these possibilities for the
electron spectrum.

\begin{enumerate}
\item If $\alpha$ steepens with decreasing frequency then the total
  energy in electrons radiating between $\nu_1$ and $\nu_2$ will
  increase as there are more particles present.  This means that $k/f$
  will be lower than we have calculated here.  This makes sense as a
  $k/f$ which is high means that the particles observed (and inferred)
  from the synchrotron radiation are not enough to create pressure
  balance with the surrounding thermal gas.  If $\alpha$ were to rise
  at low frequencies then we would ``find'' more particles and hence
  need fewer unobserved particles to obtain pressure balance.
\item This is what has been assumed in the calculations.
\item \label{flatalph}If $\alpha$ flattens out at lower frequencies, then the energy
  in the synchrotron emitting particles would be less, and so $k/f$
  would rise as from the calculated limit on $k/f$ as there would be
  the necessity for more unobserved particles to be present in the
  bubble in order to maintain pressure balance.
\item If there was a cut-off in $\alpha$ then this would cause $k/f$ to
  rise, but to an even greater extent than in situation \ref{flatalph}, as
  there are now \emph{no} low energy particles, rather than just fewer
  than before.  Our calculations assume a low-energy cut-off
  of $10\mhz$, this case is for a cut-off at higher energies than this.
\end{enumerate}

Using the spectral energy distributions from the NASA/IPAC
Extragalactic Database the spectral indices for most of the sources
remain constant, or flatten slightly at lower frequencies.  None were
seen to steepen, though the data points may be for the entire source, and
not just the extended emission.  Most sources were detected down to $100\mhz$, and
some down to $10\mhz$, so there does not appear to be a low-energy
cut-off before $10\mhz$.    No data-points were found at lower
frequencies, so the shape of the very low-energy electron population
is still unclear; however the model used to quantify the energy
present in synchrotron emitting particles is likely to be correct over
the range it has been applied.

\subsubsection{Re-acceleration}

As was discussed at length in \citet{DunnFabian04} the effect of any
re-acceleration in the bubbles could be large and difficult to
quantify.  However, the observation that, in the case of some of the
bubbles, the radio emission does not completely fill the decrement in
the X-ray emission, or that the strength of the radio emission falls
off towards the edge of the lobes leads to the conclusion that there
is little re-acceleration in these bubbles, for example Perseus
South and A2052 South.  In the Centaurus Cluster the spectral index
steepens towards the edge of the bubbles \citep{ACF_JSS_Cent05,TaylorCent02}, which also implies that
there is little re-acceleration as the electrons appear to have aged.

\section{Ghost Bubbles}\label{ghostbubbles}

\begin{figure} \centering

\includegraphics[width=1.0 \columnwidth]{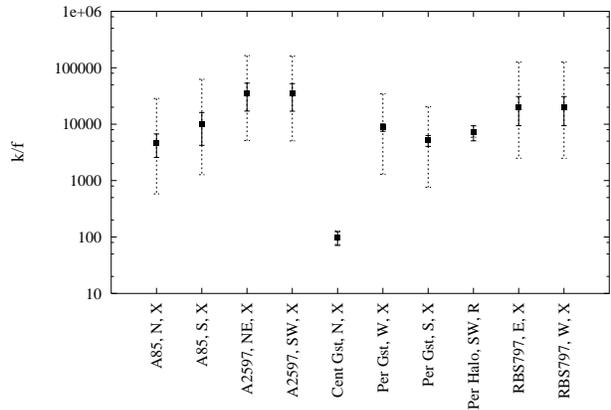}

\caption  {\label{Clustererrorghost} \small{The value of $k/f$ calculated
    from the sound speed limit for each cluster
    analysed, along with the uncertainties arising from the
    uncertainties in $\alpha$ (the dotted bars) and from the
    uncertainties in the other physical parameters of the source (the
    solid bars). }}

\end{figure}
%, though in the case of RBS797
%the input parameters for both of the bubbles are identical as the
%resolution of the data is insufficient to pick out differences

The limits on $k/f$ for the ghost bubbles can be lower limits as, if
there is no active $\ghz$ emission from the bubbles, the synchrotron
cooling time must be shorter than the age of the bubbles.  In the
cases where low frequency radio emission has been used and there is some emission from the bubbles, the limit obtained
on $k/f$ is still an upper one.  With some of these bubbles there is
insufficient radio data, and so the spectral index of the emission has
been estimated to fit in with aged sources.

\begin{figure} \centering

\includegraphics[width=1.0 \columnwidth]{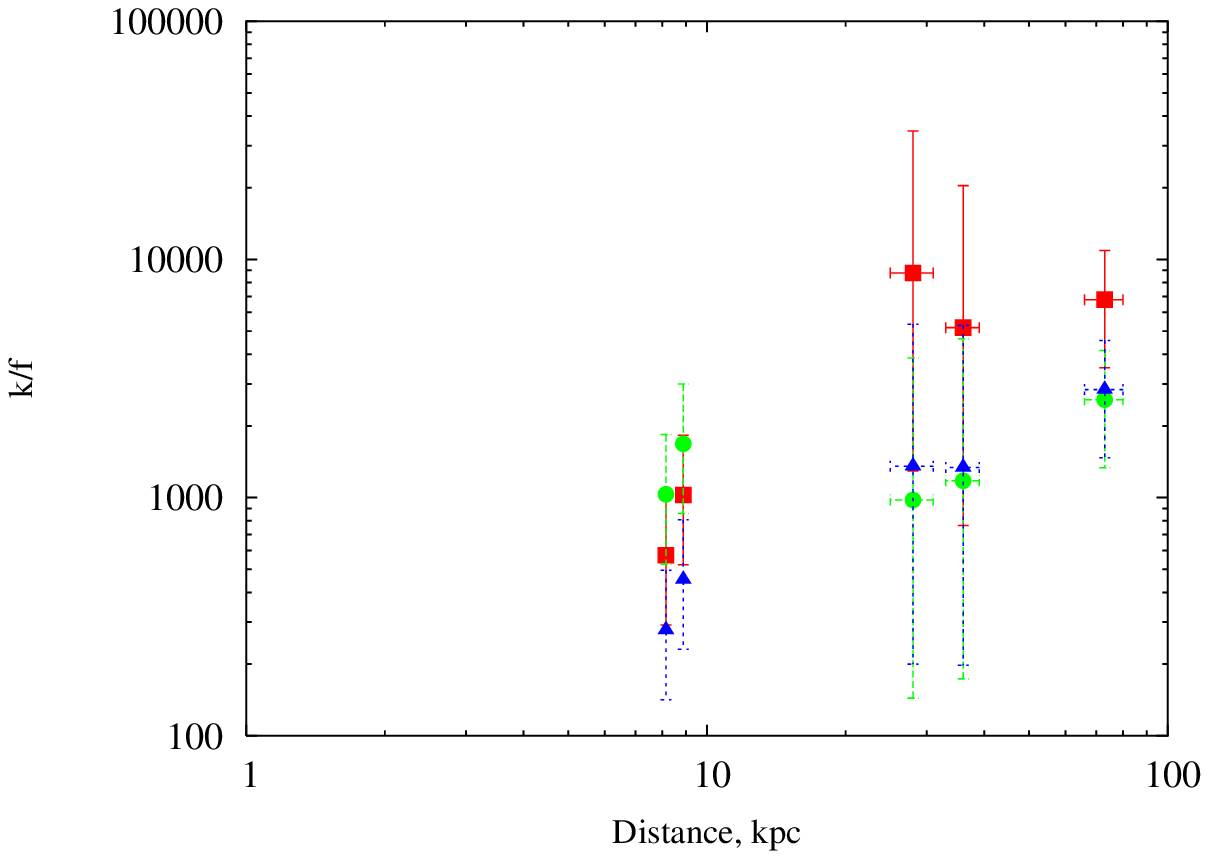}
\includegraphics[width=1.0 \columnwidth]{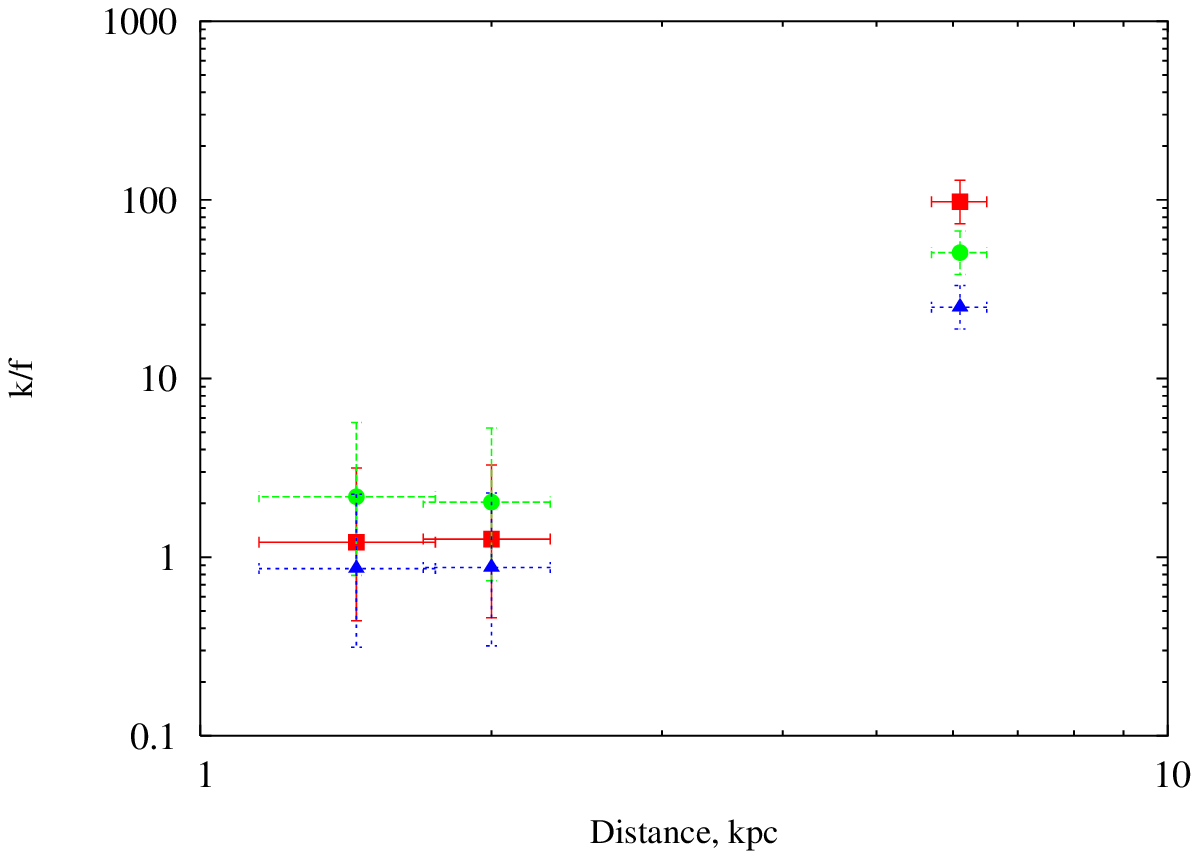}
\caption  {\label{Ghost} \small{{\scshape Top}: The change in $k/f$ for bubbles
    in the Perseus Cluster with distance from the cluster centre.
    {\scshape Bottom} The change in $k/f$ for bubbles in the Centaurus
    Cluster with distance from the cluster centre.  Estimates arising from sound-speed,
    buoyancy and refilling timescales are shown by solid squares (red),
    solid circles (green) and solid triangles (blue) respectively.
    The results have been plotted against the distance of the bubble
    from the cluster core, rather than the calculated age of the
    bubble, as the sound speed timescale for these older bubbles can be
  less than for the younger ones.  For further discussion see text.}}

\end{figure}

What can be seen from Fig. \ref{Clustererrorghost} is that all the
limits are much higher than those obtained for active bubbles.  This
means that the oldest bubbles require more unobserved particles for pressure support than the
youngest ones.  Fig. \ref{Ghost} shows more clearly the change in $k/f$ with bubble
age.  The limits have been plotted against distance from the cluster centre
rather than a timescale because of the differences in assumptions
between the types of timescale (see Section \ref{timescales}).  Fig.
\ref{Ghost} (top) and Fig. \ref{Ghost} (bottom) show the $k/f$ limits for the Perseus
and Centaurus Cluster respectively.  \citet{DunnFabian04} stated that they did
not believe that there was any dependence of $k/f$ on any physical
parameter of the cluster or radio source.  However, when looking at a
number of bubbles in a single cluster this appears not to be the
case; $k/f$ appears to rise as the bubbles age.

As the electrons in the bubbles age, the energy present in the
bubble inferred from the observed synchrotron emission decreases.
So, in order to maintain pressure balance, $k$ must rise and so $k/f$
would also be expected to rise, which is what is observed.  This implies
that there is little ongoing reacceleration in the bubble.  There may
be an effect on $f$ from the change in shape of the bubble
from a spheroid to a very oblate -- spherical-cap-like shape.  We have
taken the shapes of the Ghost Bubbles to be as oblate spheroids when
calculating their volumes and distances from the centre of the cluster
which should account for most of this change.

The pressure of the ICM falls as the bubbles rise up through it.
Therefore it could be expected that, in the absence of any change in
$f$ or the electron population, that $k/f$ would fall as fewer
particles would be required to maintain pressure balance.  However the
drop in the number of particles inferred from the synchrotron
emission, even when there is low-frequency radio data, has a greater
effect.  When there is no low-frequency radio data then many fewer
particles are inferred and this will dominate over the effect of the
drop in pressure which is of order a factor of two.

\section{Clusters with Bubbles}\label{bubblenumber}

In order to obtain a limit on the fraction of clusters which contain
bubbles the Brightest 55 (B55) sample studied by \citet{Peres1998} has been used.  The B55 sample is a $2-10 \kev$
flux-limited sample of X-ray clusters which are all close enough to
have been imaged with sufficient resolution by previous X-ray
instruments (e.g. {\it ROSAT}) and is nearly complete.

For the clusters in the sample we determined
those which  have a short cooling time
and a large temperature drop to the centre of the cluster, and as such
require some form of heating to avoid having a cooling flow.  We convert
the cooling times listed in \citet{Peres1998} to our cosmology and
take those which have a cooling time of $\leq3\gyr$ and a temperature
drop to the centre ($T_{\rm centre}/T_{\rm outer}$) of greater than a
factor of two.  There are 55 clusters in the sample, of which 23 have a $t_{\rm
  cool}\leq3\gyr$, and 19 have a large temperature drop.  For five
clusters there was no data on the cooling times in \citet{Peres1998}
and these have not been included here.  There are 17 clusters
which have both,
of which 12 have had depressions identified in the X-ray
emission.  Therefore at least 70 per cent of clusters which need some
form of heating at the centre host radio bubbles.  There are another
three clusters which contain clear bubbles whose cooling times and temperature drops miss the
cut-offs.

Out of the remaining five requiring some form of heating, one (2A 0335+096) has a
complicated X-ray core \citep{Mazzotta_0335_03}, which could conceal the presence of bubbles as
radio emission has been detected from this cluster; another (PKS
0745-191) has an amorphous radio source at the centre
\citep{Hicks_0745_02}, which may be under-pressured.  The central
nucleus is also very weak.  Another two clusters have detected radio
sources at their centres, A496 from NVSS and A2204 has an extended
source and as such may have bubbles \citep{Sanders_A2204}, but lack the
X-ray/radio resolution to see them clearly.  AWM7 has no central radio
source (as detected by NVSS), but \citet{Furusho04} find two high
metallicity blobs which could be produced by bubbles lifting up
high metallicity material from the centre of the cluster.
\citet{BurnsWhite81} find radio emission from the direction of AWM7
but identify all sources as background ones and as such there are no
known radio sources associated with the cluster.

What is surprising is that some clusters which have known bubbles fall
outside of the cut-offs that we have taken.  Hydra A and M87 both have short central cooling
times however Hydra A has a small temperature drop into the centre,
from $4$ to $3 \kev$; however the temperature map in
\citet{NulsenHydraA} shows that the core is hotter than the
surrounding gas and as such there could be a drop of greater than a factor
of two.  M87 has only a slightly larger drop
($2.5$ to $1.5 \kev$, \citet{FormanM87}), just missing the cut-off; however
{\it XMM-NEWTON} observations give a temperature drop of $2.8$ to
$1.3\kev$ \citep{BohringerM87} and so it could fall within the cut-off.  They may have
recently gone through an outburst and as such the central gas has been
heated by the radio source and has not had a chance to cool again.  MKW3s only just makes the
cooling time cut-off, however the temperature structure at the centre
of this cluster is complex \citep{MazzottaMKW04}, and could easily contain a temperature drop
of greater than factor of two.  This means that there could be 20
clusters out of the B55 sample which require some
form of heating; 15 (75 per cent) of which host bubbles.

The above results match the proportion of clusters containing radio sources presented in \citet{Eilek_conf}; who,
using the B55 sample with $\dot M>30M_\odot\yr$ and
matching these clusters to the NVSS \citet{CondonNVSS}, finds 25 out of 30
clusters (83 per cent) have a currently active radio source in the
centre.  \citet{Markovic_conf} using
the {\it ROSAT} All Sky Survey clusters which contain a massive galaxy
at the centre, coincident with the X-ray peak, and performing VLA
follow-up observations find that all 22 clusters have a currently
active central radio source.

In the recent survey of clusters by \citet{Birzan04}, out of a set
of around 80 clusters taken from the {\it Chandra} archive, they only
found 16 which contained clear bubbles.  Without discriminating between
cooling and non-cooling clusters, only 20 per cent of
clusters contained bubbles at any one time; interpreted as a duty cycle of 20 per cent.  Our
analysis shows that out of the clusters with short $t_{\rm cool}$, and as such
appear to need some form of heating at the centre, at least 70, if not 75, per cent
host radio bubbles.  Out of the total B55 sample of clusters only 30 per cent of the
clusters host bubbles, similar to the fraction presented in
\citet{Birzan04}, but those clusters which contain bubbles are likely
to have a central radio source with duty cycle of close to 100 per
cent.  The full list of which clusters have a short cooling time, large
central temperature drops and radio sources or bubbles is
shown in Table \ref{heating}.  Clusters which exhibit the cooling flow
problem commonly have an active central radio source, blowing bubbles
in the surrounding medium.  Further work is required to test whether
such activity is sufficient to stem radiative cooling or if bouts of
greater activity or some other processes are required.

\begin{table} \centering \caption{\scshape Heated Fraction}
\begin{tabular} {ccccc}
\hline
\hline
\multicolumn{3}{c}{$t_{\rm cool}\leq3\gyr$ and $T_{\rm c}/T_{\rm out}\!<\!\frac{1}{2}$}&\multicolumn{2}{c}{$t_{\rm
    cool}\!>\!3\gyr$ or $T_{\rm c}/T_{\rm out}\!>\!\frac{1}{2}$}\\
R \& B &R, no B& no R, no B & R \& B&R, no B\\
\hline
A85&2A 0335+096&AWM 7&Hydra A&3C129.1\\
A262&A496&&M87&A399\\
A426&A2204&&MKW 3s&A401\\
A478&PKS 0745-191&&&A576\\
A1795&&&&A754\\
A2029&&&&A1644\\
A2052&&&&A1650\\
A2199&&&&A3112\\
A2597&&&&A3391\\
A4059&&&&Klem44\\
Centaurus&&&&\\
Cygnus A&&&&\\
\hline

12&4&1&3&10\\
\hline
\end{tabular} \label{heating}
\begin{quote}
{\scshape Notes:} B = Bubble; R = Radio emission (may just be NVSS
source).
\end{quote}
\end{table}

\section{Conclusions}\label{concls}

Extending the sample of low-redshift clusters with clear decrements in
the X-ray emission presented in \citet{DunnFabian04}, we obtained
limits on $k/f$ , where $k$ is the ratio of the total relativistic
particle energy to that in electrons radiating between $10 \mhz$ and
$10 \ghz$, and $f$ is the volume filling fraction of the relativistic
plasma.  For all of the bubbles analysed there cannot be simple
equipartition between the particles and the magnetic field present in
the plasma.  The combined sample showed no strong dependence of $k/f$
with any physical parameter of the source -- the Rotation Measure of
the radio lobes shows some correlation but not all of the bubbles have
measured values for the RM.  In the next few years is should be
possible to measure the RMs from sources behind the cluster using
EVLA.  The correlation of the upper limits on $k/f$ with magnetic
field would then be an interesting possibility.

In the combined sample there is now no
reason to prefer two populations over one, and because of the large
spread in the values obtained on $k/f$ no conclusion about the
underlying population distribution can be made.

It was possible to trace the evolution of $k/f$ within a single
cluster in the cases of Perseus and Centaurus.  The older a bubble is
the larger its value of $k/f$ which is mainly as the result of the
aging of the relativistic electrons.

A simple study of the Brightest 55 cluster sample showed that 17,
possibly 20, clusters required some form of heating ($t_{\rm
  cool}\leq3\gyr$ and $T_{\rm centre}/T_{\rm outer}\!<1\!/2$).  Of these at least
12 (70 per cent), possibly 15 (75 per cent), contained bubbles.

\section*{Acknowledgements}

We thank Steve Allen for help with the statistical analysis, Roderick
Johnstone for computing help and the referee for helpful comments and suggestions.
ACF and RJHD acknowledge support from The Royal Society
and PPARC respectively.  GBT acknowledges support for this work from the
National Aeronautics and Space Administration through {\it Chandra} Award
Numbers GO4-5134X and GO4-5135X issued by the {\it Chandra} X-ray
Observatory Center, which is operated by the Smithsonian Astrophysical
Observatory for and on behalf of the National Aeronautics and Space
Administration under contract NAS8-03060.

\bibliographystyle{mnras} 
\bibliography{mn-jour,dunn}

\section*{Appendix} \label{appendix}\nonumber

The bubble timescales and derived powers are
presented in Table \ref{lum_t_table}.

\begin{table*} \centering \caption{\scshape Timescales and Powers}

\begin{tabular}{l l l l l l l l l l}
\hline
\hline
Cluster& Lobe$^{(1)}$&$t_{\rm CMB}^{(2)} $& $t_{\rm sound}$& $t_{\rm buoy}$
&$t_{\rm refill}$ & $\mathcal{P}_{\rm sound}^{(3)}$  &
$\mathcal{P}_{\rm buoy}$ & $\mathcal{P}_{\rm refill}$& $B_{\rm sound}$\\ 

& &$10^7 \rm{yr}$&$10^7 \rm{yr}$ &$10^7 {\rm yr}$ &$10^7 {\rm yr}$&$10^{43}{\rm ergs}^{-1}$
&$10^{43}{\rm ergs}^{-1}$ &$10^{43}{\rm ergs}^{-1}$& $10^{-5}G$\\ 

\hline
\multicolumn{9}{c}{Active Bubbles}\\
\hline

3C401  &N ,R   &8.445 &5.57 &1.58  &4.52  &1.91  &6.73  &2.36 & 0.62\\
       &S ,R   &8.445 &5.82 &2.03  &4.27  &1.33  &3.82  &1.81 & 0.60\\
4C55.16&N ,R   & 7.64 & 2.89& 2.99 &  5.18& 15.8 & 15.3 &8.81 & 9.57\\
       &S ,R   & 7.64 & 4.32& 3.00 &  5.80& 14.5 & 20.8 &10.8 & 7.32\\
A262   &E, R   & 13.9 & 1.94& 0.48 &  1.27& 0.065& 0.26 &0.10 & 1.25\\
       &W, R   & 13.9 & 1.87& 0.36 &  1.15& 0.070& 0.36 &0.11 & 1.28\\
A478   &NE, R  & 11.4 & 0.67& 0.078&  0.29& 0.44 & 3.77 &1.00 & 2.53\\
       &SW, R  & 11.4 & 1.04& 0.15 &  0.56& 1.04 & 7.25 &1.92 & 1.90\\
A1795  &NW, R  & 12.2 & 1.08& 5.13 &  1.90& 0.82 & 1.71 &0.46 & 1.85\\
       &S, R   & 12.2 & 1.26& 6.88 &  2.00& 0.67 & 1.22 &0.42 & 1.67\\
A2029  &NW, R  & 11.7 & 1.55& 1.49 &  2.89& 0.90 & 0.94 &0.49 & 1.45\\
       &SE, R  & 11.7 & 1.40& 1.57 &  3.12& 1.61 & 1.43 &0.72 & 1.55\\
M87    &E-CJ, R& 14.4 & 0.58& 0.15 &  0.59& 0.22 & 0.83 &0.21 & 2.79\\
NGC4472&E, R   &14.5  &0.99 &0.63  &1.11  &0.034 &0.053 &0.030  & 1.95\\
       &W, R   &14.5  &1.03 &0.39  &0.90  &0.049 &0.13  &0.056  & 1.90\\
NGC4636&NE, R  &14.5  &0.30 &0.15  &0.39  &0.0017&0.0033&0.0013 & 4.33\\
       &SW, R  &14.5  &0.34 &0.13  &0.36  &0.0010&0.0026&0.0010 & 3.95\\
\hline
\multicolumn{9}{c}{Ghost Bubbles}\\
\hline
A85           &N, X  & 12.4 &1.32 &0.89&1.39& 4.47& 6.62&  4.23  &1.61\\
              &S, X  & 12.4 &1.59 &1.63&1.95& 4.90& 4.80&  4.00  &1.42\\
A2597         &NE, X  & 11.5 &2.27 &2.05&2.87& 3.17& 3.50&  2.50 &1.13\\
              &SW, X  & 11.5 &3.41 &2.12&3.33& 3.38& 5.44&  3.47 &0.86\\
Centaurus     &N,X    & 13.9 &1.23 &2.48&5.10& 0.16&0.079& 0.038 &1.69\\
Perseus Ghost &W, X   & 13.9 &0.80 &7.53&5.42& 11.8& 1.25&  1.73 &2.26\\
              &S, X   & 13.9 &1.59 &7.15&6.27& 8.16& 1.81&  2.07 &1.43\\
Perseus Halo  &SW, R  & 13.9 &3.78 &10.0&9.07& 12.2& 4.60&  5.07 &0.80\\
RBS797        &W, X   & 5.90 &4.31 &4.46&8.41& 214 & 207 &  110  &0.73\\
              &E, X   & 5.90 &4.31 &4.46&8.41& 214 & 207 &  110  &0.73\\
\hline

\end{tabular} \label{lum_t_table}
\begin{quote}
{\scshape Notes:}

(1) The codes for the Lobes are N---Northern, S---Southern, E---Eastern,
    W---Western etc. , X---sizes from X-ray image, R---sizes from
    Radio image.  CJ---Counter Jet cavity in M87.

(2) The timescale for the bubble calculated from the magnetic field
    which produces the same energy density as that of the CMB at the
    redshift of the cluster for electrons radiating at $1\ghz$.

(3) The power is the PV/t work only, with $\frac{\gamma}{\gamma-1}$ not
    accounted for.  Therefore, for a fully relativistic plasma the
    values for the powers need to be multiplied by four, and for a
    non-relativistic plasma, by $5/2$.
\end{quote}
\end{table*}

\end{document}